\documentclass[11pt,a4paper]{article}

\usepackage[T1]{fontenc}
\usepackage{lmodern}
\usepackage[margin=1in]{geometry}
\usepackage{graphicx}
\usepackage{microtype}
\usepackage{hyperref}

\hypersetup{
  colorlinks=true,
  linkcolor=blue,
  citecolor=blue,
  urlcolor=blue,
  pdftitle={Celebi: Design and Realization of a Reproducible Physics Analysis Framework},
  pdfauthor={Mingrui Zhao}
}

\newcommand{\ccomment}[1]{}

\title{Celebi: Design and Realization of a Reproducible Physics Analysis Framework}
\author{%
  Mingrui Zhao\thanks{%
    \href{mailto:mingrui.zhao@mail.labz0.org}{mingrui.zhao@mail.labz0.org};
    \href{https://orcid.org/0000-0002-2858-2167}{ORCID: 0000-0002-2858-2167}.}\\
  \normalsize School of Physics, State Key Laboratory of Nuclear Physics and Technology\\
  \normalsize Peking University, Beijing, China
}
\date{}

\begin{document}
\maketitle

\begin{abstract}
Reproducibility in modern physics analysis requires more than preserving
source code or individual workflow descriptions. An analysis evolves
continuously through changes in code, configuration, parameters, input
datasets, and dependency structures, making it necessary to maintain a
coherent representation of both the analysis itself and the provenance of
its results. We present Celebi, a framework for organizing and preserving
evolving physics analyses while keeping the analysis definition lightweight
and separate from generated data products. At the system level, Celebi
maintains the analysis definition and its preserved states in an Analysis
Repository, while execution and generated results are handled separately
through a Production Factory. The middleware layer, Yuki, connects preserved
analysis states to the execution infrastructure by constructing executable
Snakemake workflows and dispatching them to supported workflow runners. This
separation allows the analysis representation to remain independent of a
particular execution backend. This paper focuses on the main design concepts
introduced in Celebi. The workflow architecture provides an explicit and
modular representation of the materials constituting an analysis and their
dependencies. A structured file organization gives this model a concrete and
human-readable realization in the analysis repository, separating reusable
algorithms, task-specific configurations, parameters, documentation, and
dependencies into well-defined objects. Finally, the impression mechanism
preserves immutable states of these objects together with the corresponding
states of their dependencies, providing persistent identities for particular
analysis configurations and allowing generated results to be associated
unambiguously with the analysis state from which they originate. Together,
these concepts provide a coherent representation of an evolving physics
analysis and establish a foundation for reproducibility, provenance tracking,
collaboration, and long-term analysis preservation.
\end{abstract}

\section{Introduction}

\ccomment{[Motivation: reproducibility requires preserving the full computational process, not only the final data products.]}
Open and reproducible research has become an important objective of the scientific community and is increasingly encouraged by policymakers and funding agencies~\cite{nature:monya,Chen:2018drk}.
Reproducibility allows scientific results to be independently verified and revisited, while openness facilitates collaboration, reuse, and the accumulation of scientific knowledge.
Achieving these goals, however, requires more than making the final data products available.
As emphasized in Ref.~\cite{Chen:2018drk}, the software, configurations, workflows, and explanations involved in producing a result are also part of the research process and should remain accessible and interpretable.
Reproducibility therefore needs to be considered throughout the research lifecycle rather than only at the stage of final publication or preservation.

\ccomment{[Scope: this work concerns the analysis of already-prepared datasets rather than detector data taking or centralized production.]}
This requirement is particularly relevant in high-energy physics, where scientific results are obtained through long chains of computational processing.
At experiments such as those at the LHC, detector signals are first processed through centrally managed production systems before being delivered to physicists in forms suitable for further analysis.
These centralized production workflows are typically standardized and maintained by the experimental collaborations.
The focus of this work begins after this stage: we consider the individual analysis of prepared datasets, where researchers develop selections, derive intermediate quantities, perform statistical analyses, evaluate systematic uncertainties, and eventually produce physics results.

Individual physics analyses are highly iterative: code, parameters, input
datasets, analysis strategies, and dependencies may all change during
development. Consequently, an analysis is better regarded as an evolving
collection of computational materials and their relationships than as a single
static workflow. A result produced at an earlier stage may depend on a
particular combination of code, configuration, parameters, input data, and
upstream processing steps, whose exact state can become difficult to reconstruct
after the analysis has evolved.

A central requirement is therefore provenance. For a generated result, it
should remain possible to identify the analysis state from which it originated,
including the relevant code, configuration, parameters, inputs, and upstream
dependencies. Maintaining this connection is essential both for reproducing
historical results and for understanding the evolution of an analysis.

\ccomment{[Existing efforts: preservation platforms and workflow systems address important parts of the reproducibility problem.]}
The importance of analysis preservation and reproducibility has long been recognized within the high-energy physics community.
Community-wide efforts on data preservation were already discussed systematically in Ref.~\cite{Kogler:2011sn}.
The CERN Analysis Preservation portal (CAP)~\cite{Chen:2016unr} provides a centralized environment in which researchers can document and preserve information about their analyses, while REANA~\cite{reana} provides infrastructure for executing reusable and containerized research workflows.
Workflow-management systems such as Snakemake~\cite{molder2021sustainable} further provide convenient mechanisms for expressing computational dependencies and automating the execution of complex analysis pipelines.
\ccomment{[Gap: existing workflow and preservation tools do not by themselves define how a continuously evolving analysis should be represented and versioned during everyday development.]}
These systems provide important building blocks for reproducible research, but
workflow execution and analysis preservation alone address only part of the
analysis-management problem considered here. During everyday development, the
analysis itself continues to change, and the correspondence between its current
structure, its historical states, and the results generated from those states
must also be maintained. Conventional workflow descriptions are generally
designed to specify how computations should be executed, whereas source-code
version-control systems record states of files and repositories. Neither
mechanism alone necessarily provides an explicit representation of an individual
analysis object together with the exact preserved states of its dependencies and
the results associated with that state. This motivates an analysis-management
layer that complements both workflow execution and conventional version control.

\ccomment{[Proposal: introduce Celebi as an analysis-management architecture that represents an evolving analysis separately from its generated data products.]}
In this work, we introduce Celebi, an analysis-management framework designed to provide such a representation.
Celebi separates the lightweight materials that define an analysis, including code, configuration, parameters, documentation, and dependency information, from the potentially large data products generated during execution.
These materials are organized in an Analysis Repository, where their relationships and evolving states can be managed explicitly.
Generated results and execution are handled separately through a Production Factory, with the Yuki middleware providing the connection between preserved analysis states and the execution infrastructure.
The detailed implementation of this execution layer is outside the scope of this paper.

\ccomment{[Paper focus: introduce the main design concepts developed in Celebi rather than the implementation details of the execution infrastructure.]}
The focus of this paper is the design of the analysis representation itself.
We introduce three closely related aspects of this design.
First, the workflow architecture describes the analysis in terms of the materials required to define each computation and the dependencies among them.
Second, the file organization provides a concrete and human-readable realization of this representation within the Analysis Repository.
Third, the impression mechanism provides immutable identities for particular states of analysis objects together with the corresponding states of their dependencies.
These concepts allow an evolving working analysis to coexist with persistent, identifiable analysis states to which generated results can be associated unambiguously. \ccomment{[Paper organization: guide the reader from the system-level view to the three design concepts.]}
The remainder of this paper first gives an overview of the Celebi system and its major components.
It then presents the workflow architecture, its realization in the repository structure, and the impression mechanism used to preserve particular analysis states and their provenance.
\section{Overview of the Celebi system}

The Celebi system separates the representation and preservation of an analysis
from the execution of the corresponding computations. Its architecture consists
of two main components: the \emph{Analysis Repository} and the
\emph{Production Factory}, connected through the Yuki middleware.

The Analysis Repository contains the lightweight materials required to define
an analysis, including source code, configurations, parameters, documentation,
and dependency information. These materials are represented as structured
analysis objects and can evolve continuously during normal development.
Selected states of these objects can be preserved as immutable
\emph{impressions}, allowing particular analysis configurations to remain
identifiable after the working analysis has changed.

\begin{figure}[hbt]
  \centering
  \includegraphics[width=\linewidth]{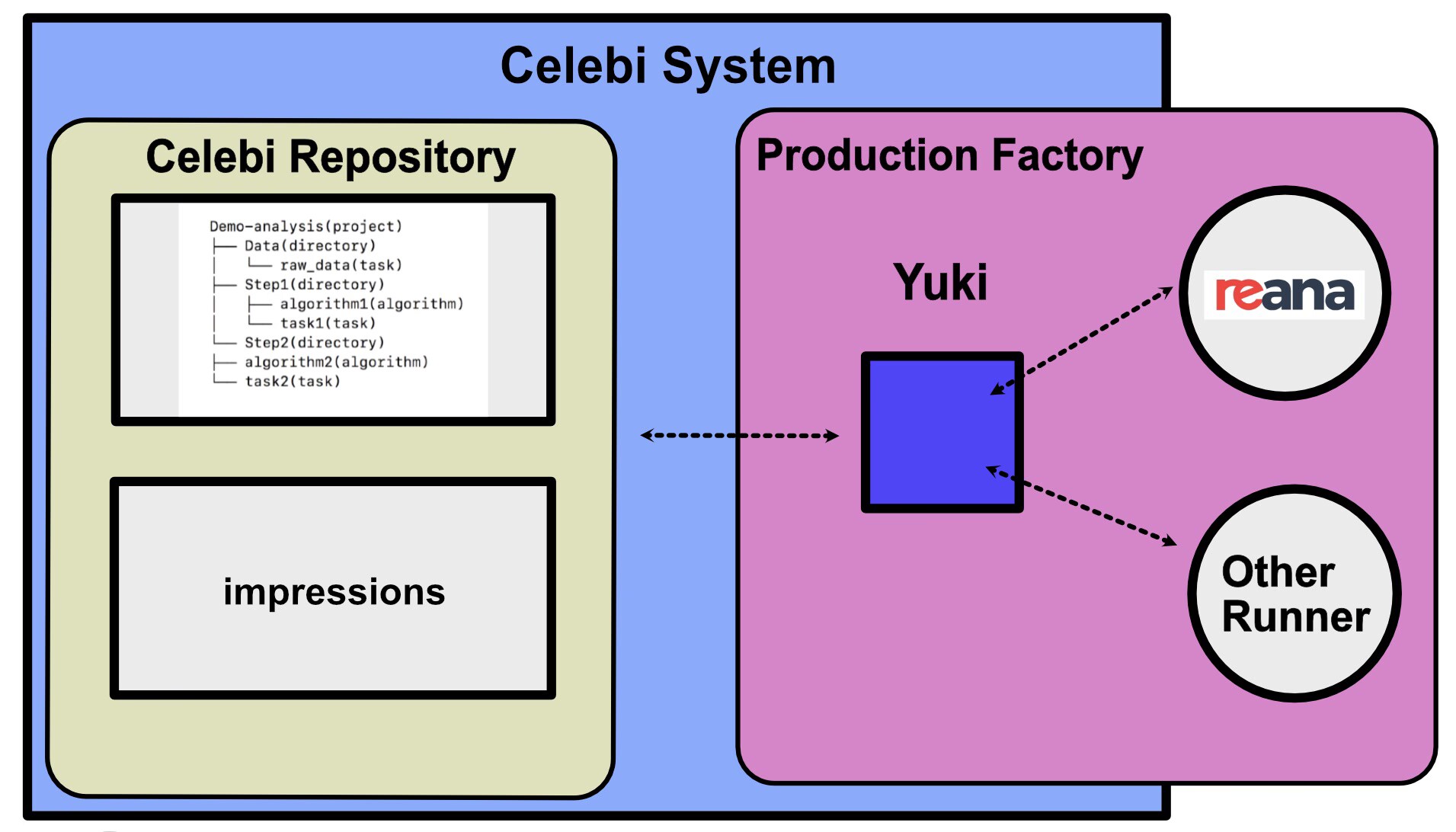}
  \caption{Overview of the Celebi system architecture. The Analysis Repository
  maintains the evolving analysis representation and its preserved states,
  while the Production Factory manages workflow execution and generated data
  products. Yuki connects the two components and dispatches executable
  workflows to supported workflow runners.}
  \label{fig:celebi-architecture}
\end{figure}

The Production Factory is responsible for executing preserved analysis states
and managing the resulting data products. This separation keeps potentially
large generated datasets outside the persistent analysis representation while
maintaining their association with the analysis states from which they were
produced. An overview of the Celebi architecture is shown in
Fig.~\ref{fig:celebi-architecture}.

Yuki provides the interface between the Analysis Repository and the execution
infrastructure. Given an impression submitted for execution, Yuki resolves its
dependency structure and constructs the corresponding Snakemake workflow. The
workflow is then dispatched to an available runner for execution. Celebi is not
tied to a particular execution backend: different runners can be used depending
on the available computing environment, while Snakemake provides a common
workflow representation across them. REANA~\cite{reana} is one currently
supported execution platform.

Yuki also monitors workflow execution and maintains the association between
the resulting outputs and the impressions from which they were produced. In
this way, the execution infrastructure can evolve independently of the
persistent analysis representation.

The following section focuses on the design of the Analysis Repository and
introduces the workflow architecture, the organization of analysis objects, and
the impression mechanism used to identify preserved analysis states.
\section{Design and realization of the Celebi analysis model}
\label{sec:design}

\ccomment{[Section overview: explain why the following three concepts are introduced together. They answer three successive questions: how an analysis is represented, how that representation is realized in the repository, and how a particular state of the evolving analysis is preserved.]}
This section introduces the main design concepts used by Celebi to represent an
evolving physics analysis. The discussion is organized around three closely
related aspects. First, the workflow architecture defines the components of the
analysis and the dependencies among them. Second, the file organization provides
a concrete and human-readable realization of this model in the Analysis
Repository. Finally, the impression mechanism preserves identifiable states of
the otherwise editable analysis. Together, these concepts connect the logical
structure of an analysis, its representation in the repository, and the
provenance of the results generated from it.
The conceptual design of the Production Factory, including Yuki, is relatively
straightforward, whereas its implementation involves substantial engineering
complexity and is therefore not discussed in detail in this paper.

\subsection{Workflow architecture}
\label{sec:workflow-architecture}

\ccomment{[Start from the conventional notion of a workflow: a DAG expresses dependencies among computational steps. This establishes the terminology before introducing what Celebi represents differently.]}
A physics analysis typically consists of a sequence of computational steps in which the output of one step may be required by another.
The resulting dependencies naturally form a directed acyclic graph (DAG), commonly referred to as a \emph{workflow}.
For a small analysis, these dependencies may be encoded implicitly in scripts, whereas more complex analyses are often expressed using dedicated workflow-description systems.
Making the workflow explicit provides a machine-readable representation of the analysis logic and facilitates automation, review, reuse, and reproducibility.

\ccomment{[Introduce the Celebi design choice at a higher level than ``materialized workflow'': Celebi represents the materials that define computations, rather than treating workflow nodes only as abstract execution steps.]}
Celebi extends this conventional view by representing not only the ordering of computations, but also the analysis materials required to define them.
Source code, computational procedures, task-specific parameters, and dependency information are treated as explicit components of the analysis representation.
The workflow therefore describes relationships among the materials that define the computation rather than only the temporal or logical ordering of execution steps.
\ccomment{[Define ``materialized workflow'' as a property of this representation, not as the complete definition of Celebi.]}
Celebi refers to this representation as a \emph{materialized workflow}.
The term \emph{materialized} emphasizes that workflow entities are represented by concrete objects in the Analysis Repository, together with the code, configuration, parameters, and dependency information required to define them.
The workflow is therefore not confined to a centralized abstract specification; it is realized directly through the analysis materials maintained in the repository.
\ccomment{[Explain one important design choice: separate the reusable computational definition from a particular use of that computation. This is the motivation for algorithm and task objects.]}
A central distinction in this model is made between an \emph{algorithm} and a \emph{task}.
An algorithm describes how a computation is performed, whereas a task describes a particular application of that computation with specified parameters and upstream dependencies.
Separating these two roles allows the same computational procedure to be reused in different parts of an analysis, or with different inputs and parameter values, without duplicating its implementation.
It also distinguishes the definition of a computational method from the particular analysis state in which that method is used.

\ccomment{[Use the Gen--Fit example to illustrate why the Celebi graph contains more information than an abstract task-ordering graph.]}
Consider a simple analysis in which a sample following a Gaussian distribution is generated and subsequently fitted.
At the level of an abstract task graph, the analysis may be represented as
$
    \texttt{Gen} \rightarrow \texttt{Fit},
$
where the edge expresses the dependency of the fitting step on the generation step.
In Celebi, the computational definitions used by \texttt{Gen} and \texttt{Fit} are represented separately from the corresponding tasks.
The \texttt{Gen} task therefore depends on its algorithm, while the \texttt{Fit} task depends on both its own algorithm and the upstream \texttt{Gen} task.
The workflow consequently records dependencies among the materials required to define the analysis, rather than only the ordering of two executable steps.
\ccomment{[Introduce the treatment of generated data: intermediate outputs belong to the task that produces them and therefore do not need to be persistent first-class workflow nodes.]}
A second design choice concerns generated intermediate data.
With the exception of externally supplied inputs, an intermediate data product is produced by a particular task and is therefore associated with that task.
Celebi consequently does not require each generated dataset to appear as an independent persistent workflow node.
Instead, a downstream task records a dependency on the upstream task responsible for producing the required data.
\ccomment{[Make the difference between data-explicit and task-to-task representations concrete.]}
For example, a data-explicit workflow could represent a dependency as
$   \texttt{task1}
    \rightarrow
    \texttt{data1.dat}
    \rightarrow
    \texttt{task2}.$
In Celebi, the persistent dependency is represented as
$   \texttt{task1}
    \rightarrow
    \texttt{task2},$
while the concrete output associated with \texttt{task1} is resolved by the
execution layer when the downstream computation is constructed.
This representation keeps potentially large generated datasets outside the persistent workflow description while retaining the dependency needed to identify their origin.

\ccomment{[Explain the provenance argument carefully without claiming strict mathematical determinism of the output. The persistent specification identifies the provenance-relevant state of the computation.]}
The persistent representation therefore records the provenance-relevant materials required to identify a computation: its algorithm, parameters, upstream dependencies, external inputs, and execution conditions.
Under controlled execution conditions, this information provides the basis for reproducing the intended computation without requiring its generated outputs to be included directly in the workflow definition.
External datasets form an important exception because they originate outside the workflow and have no upstream producing task; their representation is discussed in Sec.~\ref{sec:file-organization}.

\ccomment{[Introduce distributed dependency specification as another property of the workflow architecture. Each component owns its local dependencies; the global DAG emerges from their composition.]}
The workflow specification is distributed among the analysis objects.
Rather than requiring a single global file containing the complete DAG, each relevant object records its own direct dependencies.
The complete workflow can then be reconstructed from these local relationships.
This locality allows an analysis component and its dependencies to evolve without requiring unrelated parts of a monolithic workflow description to be rewritten.

\ccomment{[Conclude the workflow subsection by summarizing what is persistently represented. Do not introduce impressions yet.]}
The resulting workflow consists primarily of lightweight and versionable analysis materials rather than generated data products.
A particular state of these materials defines a provenance-relevant state of the analysis from which computations can be carried out.
The next subsection describes how this abstract representation is realized concretely in the structure of a Celebi repository.

% \begin{comment}
% \begin{figure}[htb!]
%    \centering
%    \includegraphics[width=0.7\textwidth]{figures/workflow_concept2}
%    \caption{
%        Example of the Celebi workflow representation.
%        Tasks and the algorithms defining their %computation are represented separately, while %generated intermediate data are associated with the %tasks that produce them.
%     }
%    \label{fig:workflowconcept2}
%\end{figure}
%\end{comment}

\subsection{File organization}
\label{sec:file-organization}

\ccomment{[Bridge from the abstract model to its concrete realization: the repository should make the workflow structure visible and editable using ordinary file-system concepts.]}
The workflow model described above is realized in a \emph{Celebi repository}.
The repository contains the lightweight materials required to define the analysis, including source code, configurations, parameters, dependency metadata, and documentation.
Workflow entities are represented as structured objects in the file system, so that the logical organization of the analysis has a direct and human-readable representation in the repository.

\ccomment{[Define the four object types while clarifying the distinction between organizational objects and computational objects. This avoids the ambiguity of saying that objects cannot contain other objects while project/directory are themselves containers.]}

Celebi defines four object types: project, directory, algorithm, and task. They
serve two complementary roles. Project and directory objects provide the
hierarchical organization of the repository, while algorithm and task objects
represent the materials that participate directly in the workflow.

\begin{itemize}

\item \textbf{Project}: the root object of a Celebi repository, representing
the analysis as a whole.

\item \textbf{Directory}: an organizational object used to group related
analysis components according to a human-readable hierarchy.

\item \textbf{Algorithm}: a reusable definition of a computational procedure,
including the code and configuration required to describe how the computation
is performed.

\item \textbf{Task}: a concrete workflow object describing the use of analysis
inputs within a particular part of the workflow. For a computational step, a
task associates an algorithm with the corresponding parameters and upstream
dependencies. A task may also provide the interface through which externally
produced data enter the workflow.

\end{itemize}

\ccomment{[Explain why algorithm and task are separate in the repository, rather than focusing on the names of YAML files or command syntax.]}
For computational steps, the separation between algorithms and tasks
distinguishes reusable computational logic from its concrete invocation. For
example, the same fitting algorithm may be used by several tasks corresponding
to different datasets or systematic variations. The algorithm object can remain
unchanged, while each task records the particular parameters and dependencies
relevant to that invocation. This separation reduces duplication and makes it
explicit whether a variation originates from a change in the computational
procedure or from the conditions under which it is applied.

\ccomment{[Explain how dependencies are represented without coupling the persistent analysis description to generated file paths.]}
Dependencies are expressed between objects rather than through hard-coded paths to generated outputs.
A task can assign local aliases to its dependencies, allowing its computational definition to refer to upstream analysis components through stable logical names.
For an internally generated input, the dependency therefore refers to the upstream task responsible for producing the data rather than to the physical location of a particular output file.
This realizes the task-to-task dependency model introduced in Sec.~\ref{sec:workflow-architecture} and separates the persistent analysis structure from transient data locations.

\ccomment{[Explain external inputs at the conceptual level. They have no producing task inside the workflow, so Celebi represents them through a source-like task to preserve a uniform downstream dependency model.]}
External input datasets require a slightly different treatment because they are
not generated by an upstream computation within the analysis. Celebi represents
such an input through a task object that records how the externally supplied
data are identified and exposes them to the workflow as an available output.
Such a task therefore defines the boundary at which data produced outside the
Celebi workflow enter the analysis, rather than representing a computation that
produces the data itself.

Downstream tasks can depend on this object through the same task-to-task
dependency mechanism used for internally generated data. For an internally
generated input, the upstream task identifies the computation responsible for
producing the required output; for an external input, the corresponding task
instead identifies data that already exist outside the workflow. In this way,
both internally generated and externally supplied inputs can be incorporated
through a uniform dependency model without introducing persistent data objects
as separate workflow nodes.

\ccomment{[Explain documentation and metadata only at the level needed for the design paper; exact filenames and schemas belong to a later implementation paper.]}
Each repository object also carries local metadata and documentation describing its role and configuration.
Keeping this information with the corresponding analysis component makes the repository self-describing at the level of individual objects and supports both human inspection and programmatic interpretation of the workflow.

\ccomment{[Close the subsection by returning to the design objective: the repository contains the definition of the analysis, while large generated products remain outside it.]}
With this organization, the Analysis Repository contains the relatively small materials that define the analysis rather than its potentially much larger generated products.
The repository can therefore remain lightweight as the analysis evolves, while still retaining the information required to describe its logical structure and computational configuration.
The remaining problem is how to identify a particular state of this continuously editable representation.
Celebi addresses this through the impression mechanism described below.

\subsection{Impressions}
\label{sec:impressions}

\ccomment{[Motivate impressions from the tension between an editable working analysis and the need to refer permanently to the exact state that produced a result.]}
Objects in the Analysis Repository represent the current working state of an analysis and are expected to change during normal development.
Reproducibility, however, requires a stable way to identify a particular state even after the working analysis has evolved.
Celebi introduces \emph{impressions} to separate these two roles: objects remain editable, whereas selected states of those objects can be preserved immutably.

\ccomment{[Give the central definition: an impression preserves both the local object state and the exact dependency states on which it relies.]}
An impression is an immutable representation of a Celebi object at a particular analysis state.
It records the materials defining the object together with references to the impressions corresponding to its dependencies.
An impression therefore identifies not only the local state of one object but also the precise preserved states of the upstream objects on which it depends.

\ccomment{[Explain how dependency references recursively define a preserved workflow state.]}
Each impression is assigned a persistent identifier that remains associated
with that immutable state. For an object with dependencies, the impression
metadata records the identifiers of the exact predecessor impressions on which
it depends. These immutable references preserve the dependency state of the
workflow at the time the impression is created. Consequently, two preserved
states may differ because of a change in an upstream dependency even when the
local contents of the object itself remain unchanged.

\ccomment{[Explain immutability: editing the working object does not modify history; it creates a new state that may be preserved separately.]}
Once an impression has been created, it is not modified.
Changes to the corresponding code, parameters, configuration, or dependencies instead produce a new working state, which may subsequently be recorded as a new impression.
The repository can therefore continue to evolve without altering previously preserved analysis states.

\ccomment{[Clarify the relation to conventional source-code version control. Git preserves repository snapshots; an impression additionally expresses workflow-local identity and exact dependency-state relationships. The mechanisms are complementary.]}
The impression mechanism is complementary to conventional source-code version control.
A version-control commit records a state of a repository as a collection of
files, whereas an impression identifies the state of an individual analysis
object together with the exact preserved states of the workflow dependencies
relevant to that object. The two mechanisms therefore operate at different
levels: version control records repository history, while impressions provide
analysis-level identities that can be associated directly with generated
results. Celebi uses impressions as a complement to, rather than a replacement
for, conventional source-code version control.

\ccomment{[Connect impressions to the lightweight-repository design: generated datasets are not copied into every preserved state.]}
Because the persistent analysis representation consists primarily of lightweight materials, impressions can also remain compact.
Code, parameters, configurations, and dependency information can be preserved directly, while potentially large generated datasets remain outside the impression itself.
Instead, such datasets are associated with the task impression under which they were produced.

\ccomment{[State the key provenance property: a result is associated with a task impression, whose dependency closure identifies the complete preserved analysis state behind it.]}
For a task impression, the combination of its local preserved state and the impressions of its dependencies identifies the analysis configuration associated with that computation.
A generated result can therefore be bound to the corresponding task impression.
Following the dependency chain provides access to the code, parameters, configurations, and upstream input states associated with the production of that result.

\ccomment{[Explain current versus historical states conceptually, without introducing the status() API or CLI implementation.]}
When the working representation of an object changes, it no longer corresponds to its previous impression.
Results associated with that previous impression are therefore historical results rather than results of the current working state.
They are not discarded, however: they remain associated with the immutable impression under which they were produced.
Celebi can consequently distinguish the evolving current analysis from preserved historical states while maintaining the provenance of results from both.

\ccomment{[Close the whole section by connecting the three concepts: workflow architecture defines relationships, file organization realizes them, and impressions preserve identifiable states of the resulting structure.]}
The three concepts introduced in this section play complementary roles.
The workflow architecture defines the analysis components and their dependencies; the file organization realizes this representation as a structured and editable repository; and impressions provide persistent identities for selected states of that evolving structure.
Together, they allow Celebi to maintain a direct relationship between the organization of an analysis, its evolution over time, and the provenance of the results produced from it.
\section{Summary}

This work presents Celebi, a framework for representing and preserving evolving
physics analyses. Celebi organizes analysis code, configurations, parameters,
external inputs, and dependencies into a lightweight materialized workflow,
where the components defining an analysis are represented directly as structured
repository objects rather than only as abstract execution steps.

Reproducibility and provenance are supported through immutable impressions,
which preserve the state of an object together with references to the exact
preserved states of its dependencies. Generated results can therefore remain
associated with the analysis state from which they were produced even as the
working analysis continues to evolve.

Execution is separated from this persistent analysis representation. The Yuki
middleware converts preserved analysis states into Snakemake workflows and
dispatches them to supported workflow runners, allowing different execution
backends to be used without changing the underlying analysis model. REANA is one
currently supported execution platform.

Together, the materialized workflow, structured repository organization, and
impression mechanism provide a foundation for provenance tracking, reproducible
execution, collaborative development, and long-term preservation of
high-energy-physics analyses.

\bibliographystyle{unsrt}
\bibliography{references}

@article{nature:monya,
  author = {Baker, Monya},
  title = {{1,500 scientists lift the lid on reproducibility}},
  journal = {Nature},
  volume = {533},
  pages = {452--454},
  year = {2016},
  doi = {10.1038/533452a}
}

@article{Chen:2018drk,
  author = {Chen, Xiaoli and others},
  title = {{Open is not enough}},
  journal = {Nature Phys.},
  volume = {15},
  pages = {113--119},
  year = {2019},
  doi = {10.1038/s41567-018-0342-2}
}

@article{Kogler:2011sn,
  author = {Kogler, Roman and South, David M. and Steder, Michael},
  title = {{Data Preservation in High Energy Physics}},
  journal = {J. Phys. Conf. Ser.},
  volume = {368},
  pages = {012026},
  year = {2012},
  doi = {10.1088/1742-6596/368/1/012026}
}

@article{Chen:2016unr,
  author = {Chen, Xiaoli and others},
  title = {{CERN Analysis Preservation: A Novel Digital Library Service
            to Enable Reusable and Reproducible Research}},
  journal = {Lect. Notes Comput. Sci.},
  volume = {9819},
  pages = {347--356},
  year = {2016},
  doi = {10.1007/978-3-319-43997-6_27}
}

@article{reana,
  author = {{\v{S}}imko, Tibor and others},
  title = {{REANA: A System for Reusable Research Data Analyses}},
  journal = {EPJ Web Conf.},
  volume = {214},
  pages = {06034},
  year = {2019},
  doi = {10.1051/epjconf/201921406034}
}

@article{molder2021sustainable,
  author = {M{\"o}lder, Felix and others},
  title = {{Sustainable data analysis with Snakemake}},
  journal = {F1000Research},
  volume = {10},
  pages = {33},
  year = {2025},
  doi = {10.12688/f1000research.29032.3}
}

\end{document}